\documentstyle[preprint,aps]{revtex}
\begin{document}
\draft
\preprint{}


\title{Magnon-magnon interactions in the Spin-Peierls compound CuGeO$_3$}
\author{Claudius Gros$^1$, Wolfgang Wenzel$^1$,
        Andreas Fledderjohann$^2$\\
        P. Lemmens$^3$, M. Fischer$^3$, G. G\"untherodt$^3$,
        M. Weiden$^4$, C. Geibel$^4$, F. Steglich$^4$
       }
\address{$^1$ Institut f\"ur Physik, Universit\"at Dortmund,
         44221 Dortmund, Germany\\
        }
\address{$^2$ Physics Department, University of Wuppertal,
         42097 Wuppertal, Germany\\
        }
\address{$^3$ 2. Physikalisches Institut, RWTH Aachen,
         52056 Aachen, Germany\\
        }
\address{$^4$ FB Technische Physik, TH-Darmstadt, 
         Hochschulstr. 8, 64289 Darmstadt, Germany\\
        }
%
\date{\today}
\maketitle
\begin{abstract}
In a magnetic substance the gap in the Raman spectrum, $\Delta_R$,
and the neutron scattering gap, $\Delta_S$,
are related by $\Delta_R/\Delta_S\approx2$
if the the magnetic excitations (magnons) 
are only weakly interacting.
But for CuGeO$_3$ the experimentally observed 
ratio is of the order
$\Delta_R/\Delta_S\sim1.49-1.78$,
indicating attractive magnon-magnon interactions
in the quasi-1D Spin-Peierls compound CuGeO$_3$.
We present numerical estimates of
$\Delta_R/\Delta_S$ from
exact diagonalization studies for finite chains and
find agreement with experiment for intermediate
values of the frustration parameter $\alpha$.
An analysis of the numerical Raman intensity
leads us to postulate a {\it continuum} of
two-magnon bound states in the Spin-Peierls
phase.  We discuss in detail the numerical method
used, the dependence of the results on
the model parameters and a novel matrix-element
effect due to the dimerization of the Raman-operator
in the Spin-Peierls phase.

\end{abstract}
The Spin-Peierls compound CuGeO$_3$ has
lately been studied intensively \cite{Hase}
and found to exhibit well defined
magnetic excitations (magnons) in
the dimerized phase \cite{Nishi}.
These magnons are found \cite{Ain} to be seperated
by a gap from the continuum of two-spinon
excitations predicted for the
Heisenberg-chain \cite{Mueller} and were
recently observed in KCuF$_3$ \cite{KCuF3} 
and in CuGe$O_3$ \cite{Arai}.
In this context it was realized 
\cite{Tsvelik,Uhrig,Fledderjohann}
that the magnons in the dimerized Heisenberg
chain can be regarded as two-spinon bound-states. While
spinons are essentially free in a homogeneous spin chain they
interact strongly in a (gapped) dimerized spin chain where
magnons are well defined excitations with dispersion $\omega_q$ 
contributing a delta-function
$\sim\delta(\omega-\omega_q)$ to
the dynamical structure factor, $S(q,\omega)$.

It is then all but natural to investigate the
interactions of two magnons in a dimerized spin-chain.
Here we will present numerical and experimental
evidence that magnons do strongly interact in dimerized
spin chains leading to a continuum of two-magnon
bound states.
For this purpuse we will present
data from exact diagonalization of chains with
up to $N_s=28$ sites and experimental Raman spectra
for CuGeO$_3$. We will, in particular, investigate
the gap $\Delta_S$ observed in $S(q,\omega)$ and
the gap $\Delta_R$ observed in the two-magnon Raman spectrum 
$I_R(\omega)$. We find generally that
$\Delta_S/\Delta_R<2$, indicating strong magnon-magnon interactions
in 1D dimerized spin systems.

As the minimal model for magnetic excitations in CuGeO$_3$
one can consider the frustrated 1D spin Hamiltonian
\begin{equation}
H = J\sum_i\, [ (1+\delta(-1)^i)\,{\bf S}_i\cdot{\bf S}_{i+1}
              +    \alpha\,       {\bf S}_i\cdot{\bf S}_{i+2}
              ]~,
\label{H}
\end{equation}
where $\delta$ is the dimerization parameter that vanishes above
$T_{SP}$ \cite{Castilla,Riera}.
The special geometry \cite{Braden,Khomskii}
of the super\-exchange
path in CuGeO$_3$ along the c-axis
leads to a small value of the
exchange integral $J\approx150$K and a substantial
n.n.n. frustration term $\sim\alpha$ which competes
with the n.n. anti\-ferromagnetic exchange.
The correct value of $\alpha$ suitable
for CuGeO$_3$ is still under discussion.
While Castilla {\it et al.} proposed
$\alpha\approx0.24$ \cite{Castilla}, a much
larger value $\alpha\approx0.35$ was
proposed by Riera and Dobry \cite{Riera}
and, recently, by Brenig {\it et al.} \cite{Brenig}.
The interchain couplings have been
estimated to be small, $J_b\approx0.1J$ and
$J_a\approx-0.01J$ for the interchain exchange
constants along a- and b- directions, respectively
\cite{Nishi}.

The phase diagram of $H$ in Eq.\ (\ref{H})
has been calculated using the 
density-matrix renormalization-group method \cite{Chitra}.
For $\delta=0$ and $\alpha<\alpha_c\approx0.2411$,
the ground state is gapless and renormalizes to the
Heisenberg fixed point.  For $\alpha=0.5$
and $\delta=0$, the ground state is given by a valence-bond
state with a gap of order $J/4$ induced by frustration.

An experimental method particularly suited for the
study of magnetic excitations in an
antiferromagnet is two-magnon
Raman scattering.
For CuGeO$_3$, the Raman operator in $A_{1g}$ 
symmetry \cite{Fleury} is proportional \cite{RC} to 
\begin{equation}
H_R = \sum_i\,(1+\gamma(-1)^i)\,{\bf S}_i\cdot{\bf S}_{i+1}~.
\label{H_R}
\end{equation}
In the homogeneous state ($\delta=\gamma=0$) 
the Raman operator 
commutes with the Heisenberg Hamiltonian
for the case $\alpha=0$ and 
there would be no Raman scattering \cite{RC,Singh}.
However when $\alpha \neq 0$, the model (\ref{H}) leads to magnetic
Raman scattering $\sim\alpha^2$
due to the presence of competing interactions
which can be observed experimentally.
Note the presence 
of the factor $\gamma$ in Eq.\ (\ref{H_R}) which
appears for $T<T_{SP}$ because the exchange integral is sensitive 
to the inter-ionic distance.  


We have exactly diagonalized Eq.\ (\ref{H}) for chains
with up to 28 sites by a generalized Lanczos method 
and evaluated the Raman spectral weight at
zero temperature,
\begin{equation}
I_R(\omega) \,=\, -{1\over\pi}Im\,
\langle0|H_R{1\over \omega+i\epsilon - (H-E_0)}H_R|0\rangle,
\label{I_R}
\end{equation}
where $E_0$ is the ground state energy, $H$ the
Hamiltonian given by Eq.\ (\ref{H}) and $\epsilon\rightarrow0+$.
We have also calculated the dynamical structure factor
\begin{equation}
S(q,\omega) \,=\, \sum_n
\big|\langle n|S_{q}^z|0\rangle\big|^2\delta(\omega-(E_n-E_0))
\label{S(q,o)}
\end{equation}
where $S_q^z=N_s^{-1/2}\sum_{l=1}^{N_s}\exp[iql]S_l^z$ and
$|n\rangle,\ E_n$ are the Eigenstates and
Eigenenergies of the spin chain,
respectively.  

We have evaluated $S(q,\omega)$ for chains with
up to $N_s=24$ sites using an approximate scheme
for the determination of the low lying excitation
energies $E_n-E_0$ and the corresponding
transition probabilities $w_n(q)=|\langle n|S_q^z|0\rangle|^2$
\cite{rec}. Using a recursion algorithm a set of orthogonal
states is built starting with $S_q^z|0\rangle$.
Coefficients occuring in this procedure form a tridiagonal
matrix whose eigenvalues and eigenstates determine
the excitation energies and transition probabilities \cite{Fledderjohann,rec}.

For a numerical evaluation of Eq.\ (\ref{I_R}), we have used
the {\it kernel polynomial approximation}. Since the advantages
of this method have been realized only recently \cite{Silver}
we give here a brief account.
We start by rescaling the
Hamiltonian by $H=cX+d$ such that the eigenvalues of
the rescaled Hamiltonian $X$ are in the interval
$[-1,1]$. Similary we define a rescaled energy and frequency
by $E_0=cx_0+d$ and $\omega=cx+d$ and expand 
$I_R(x)$ in terms of Tschebycheff polynomials, 
$T_l(x)$:
\begin{equation}
I_R(x) = {1\over\sqrt{1-x^2}}\sum_{l=0}^{N_p}\, a_l\, T_l(x+x_0),
\label{expansion}
\end{equation}
where the number of polynomials retained, $N_p$, determines
the accuracy of the approximation which becomes exact 
in the limit $N_p\rightarrow\infty$. 
The expansion coefficients $a_l$ are determined using the
orthogonality relations for Tschebeycheff polynomials to be
$$
a_l={2-\delta_{l,0}\over\pi}\,\langle0|\,H_R\,T_l(X)\,H_R\,|0\rangle,
$$
which can be evaluated recursively via the formula
$T_{l+1}(x)=2xT_l(x)-T_{l-1}(x)$. The advantage of
an expansion in orthogonal polynomials is its 
numerical stability. It is indeed possible to
evaluate several thousands of $a_l$ recursively without
encountering numerical instabilities.
Often we will use only  a limited number $N_p=100$
for comparison with experimental data. 

A truncated expansion in orthogonal polynomials will,
in general, lead to unwelcome 
Gibbs oscillations  for any finite $N_p<\infty$.
These Gibbs oscillations have been studied carefully
in the past \cite{Silver} and can be suppressed
efficiently and reliable by the
replacement $a_l\rightarrow a_l g(z_l)$
with $g(z_l) = [\sin(\pi z_l)/(\pi z_l)]^3$ and
$z_l = l/(N_p+1)$ \cite{Silver}. Note that this replacement,
which we use throughout this paper,
still satisfies the correct limit $N_p\rightarrow\infty$.

A finite $N_p<\infty$ in (\ref{expansion}) does broaden the 
delta-poles in a finite system calculation. 
The dependence of $I_R(\omega)$ on
$N_p$ is illustrated in Fig.\ \ref{N_p} for 
a chain with $N_s=28$ sites, $\alpha=0.24$ and $\delta=0=\gamma$.
In the inset of Fig.\ \ref{N_p} we compare the
broading of a single pole with the kernel polynomial
approximation ($N_p=100$) with a Lorentzian
$\epsilon/((\omega-\omega_i)^2+\epsilon^2)$ and
$\epsilon=0.025$. Note the absence of the high-energy
tails in the kernel polynomial approximation.

%

We have measured the Raman intensity on single crystal
CuGeO$_3$ using
the excitation line $\lambda=514.5$-nm 
of an Ar-laser with a laser power of 2.7mW. 
We ensured that the incident radiation does not
increase the temperature of the sample by more than 1.5 K.
We used a DILOR-XY spectrometer and a nitrogen cooled CCD 
(back illuminated) as a detector in a quasi backscattering 
geometry with the polarization of incident and scattered light parallel 
to the c-axis and the Cu-O chains, respectively.


In Fig.\ \ref{T=20}, we present the data for the 
two-magnon Raman continuum in the homogeneous 
state at $T=20$ K. 
Phonon lines \cite{Raman,Lemmens} at 184cm$^{-1}$ 
and at 330cm$^{-1}$ are
subtracted from the experimental 
data (squares). The experimental Raman spectrum presented in
Fig.\ \ref{T=20} is in agreement with other Raman
studies on CuGeO$_3$ \cite{Raman}.
We have included in
Fig.\ \ref{T=20} the numerical results for
$I_R(\omega)$ obtained for chains with $N_s=24$ (dashed lines)
and $N_s=28$ (solid lines) sites for $\delta=0=\gamma$
and $N_p=100$. Note that the finite-size
effects are quite small. We show in Fig.\ \ref{T=20}
data for two parameter sets, namely
$\alpha=0.24,\ J=150K$ and $\alpha=0.35,\ J=159K$.
We note that $\alpha=0.35$, which is favoured by fits to
the susceptibility \cite{Riera} and to the specific
heat \cite{Brenig} does not agree well with the
Raman spectrum. Similar results have been obtained
previously with a solitonic mean-field approach to the frustrated
Heisenberg chain \cite{RC,susc}.


In the dimerized phase $\delta\ne0$ we have found that the
numerically obtained Raman spectrum depends very much on
the dimerization parameter $\gamma$ in the Raman operator
as we illustrate in Fig.\ \ref{gamma} for a chain with
$N_s=24$ sites, $\delta=0.03,\ \alpha=0.24$ and $N_p=100$.
Between $\gamma=0$ and $\gamma=0.15$ the spectrum changes
qualitatively and a low-energy peak can be resolved
for $\gamma=0.12,0.15$, but not for $\gamma=0$.

In order to understand the dramatic dependence of the
$I_R(\omega)$ on $\gamma$ observed in Fig.\ \ref{gamma}
we have analyzed the dependence of the seven
lowest poles contributing to $I_R(\omega)$ on $\gamma$,
illustrated in the inset of
Fig.\ \ref{weight_gamma} for $N_s=24$, $\alpha=0.24$,
$\delta=0$, $N_p=1000$ and $\gamma=0$.

We start by rewriting Eq.\ (\ref{I_R}) in the form
\begin{equation}
I_R(\omega) \,=\, \sum_n
\big|\langle 0|H_R|n\rangle\big|^2\delta(\omega-(E_n-E_0))
\label{I_new}
\end{equation}
and noting that we can decompose the Raman operator
into two parts, $H_R=H^\prime+\gamma H^{\prime\prime}$.
The weight of an excited state then becomes
\begin{equation}
w_n =
\big|\langle 0| H^\prime+\gamma H^{\prime\prime} |n\rangle \big|^2
= \big|m^\prime+\gamma m^{\prime\prime}\big|^2,
\label{w_n}
\end{equation}
and for any $n$ there is a $\gamma_0=-m^\prime/m^{\prime\prime}$
for which $w_n$ vanishes.
We have analyzed the energy-dependence of $\gamma_0=\gamma_0(E_n)$
and found that for the five dominant poles,
$n=1,2,4,6,7$
\begin{equation}
\gamma_0\ \approx\ \delta + {4\over 3000}\, E_n,
\label{gamma_0}
\end{equation}
in inverse wavelength $[cm^{-1}]$ for $E_n$.
This dependence of $w_n$ on $E_n$ is shown
in Fig.\ \ref{weight_gamma} where we have plotted
the weights as a function of $\gamma-\gamma_0(E_n)$.
The data presented in Fig.\ \ref{weight_gamma} 
clearly indicates that the dominant contributions
to $I_R(\omega)$ follow the scaling
relation
\begin{equation}
\rho(\gamma,E_i) = {I_{\gamma=0}\over I_\gamma}
\left({\gamma-\gamma_0(E_i)\over\gamma_0(E_i)}\right)^2,
\label{erasor}
\end{equation}
where $I_\gamma$ is a normalization constant, which we
have approximated by the constraint
$\int_0^{\omega_c}d\omega\rho(\gamma,\omega)=1$,
with $\omega_c=6J$. The scaling relation
Eq.\ (\ref{erasor}) constitutes, on the other hand,
also the rescaling of the Raman intensity 
$I_R(\omega)=I_R(\gamma,\omega)$
with $\gamma$, such that we can write
\begin{equation}
I_R(\gamma,\omega)\approx\rho(\gamma,\omega)I_R(0,\omega).
\label{analytic}
\end{equation}
We see that the effect of $\gamma$ is to
``burn a spectral hole'' in $I_R(0,\omega)$ at a characteristic
frequency $\gamma_0(\omega)$, in agreement with the numerical
results presented in Fig.\ \ref{gamma}.

We can actually find a complete analytic formula for
$I_R(\gamma,\omega)$ by noting that for $\alpha=0.24$, $\delta=0.03$
we can approximate $I_R(0,\omega)$ 
by the expression (see Fig.\ \ref{gamma}
and \cite{RC})
\begin{equation}
I_R(\gamma=0,\omega) = A\theta(\omega-\Delta_R)
\left(1-\tanh[2(\omega-\omega_0)]\right),
\label{I_0}
\end{equation}
with the values of the parameters
$\Delta_R\approx30{\rm cm}^{-1}$,
$\omega_0\approx312{\rm cm}^{-1}$ being
determined by a fit to the numerical data
(A is a normalization constant). The heavy-side
function
$\theta(\omega-\Delta_R)$ in Eq.\ (\ref{I_0})
reflects the absence of two-magnon excitations
below $\Delta_R$ in the dimerized state.


In Fig.\ \ref{T=5}, we present the experimental
Raman spectrum in the spin-Peierls phase at $T=5$K
(with phonon lines substracted, filled squares) 
in comparision with the analytic result
for $\gamma=0.12$(Eq.\ (\ref{analytic}) together with
Eq.\ (\ref{I_0}), solid line).
Experimentally a two-magnon peak is observed at
30cm$^{-1}$ and a broad continuum at higher
frequencies. These two features are well reproduced
by the analyitic result.
There is, on the other hand, a new, probably magnetic
line at 225cm$^{-1}$, which is not
reproduced by our theoretical study of the one-dimensional spin
model (\ref{H}), compare also Fig.\ \ref{gamma}. This
fact has led us to speculate \cite{RC} that the
interchain coupling $J_B$ might become relevant
for $T<T_{SP}$ in CuGeO$_3$. 


Which is the physical significance of the peak
at $\Delta_R\sim30{\rm cm}^{-1}$  in Fig.\ \ref{T=5}
observed both in experiment and in our analytical result?
Neutron-scattering experiments \cite{Nishi,Martin} indicate 
a one-magnon gap $\Delta_S\sim(2.1-2.5)$meV,
i.e.\ $\Delta_S\sim(16.9-20.2){\rm cm}^{-1}$.
The experimental ratio
\begin{equation}
{\Delta_R\over\Delta_S}\bigg|_{\rm Exp.}\ \sim\ 1.49-1.78
\label{ratio_exp}
\end{equation}
is smaller than two, the value expected for non-interacting
magnons. This indicates an attractive interaction between
magnons in 1D dimerized spin chains, similar to the attraction
between spinons. Uhrig and Schulz \cite{Uhrig} have indeed 
postulated an isolated singlet bound state 
with energy $E_s$ and $2\Delta_S>E_s>\Delta_s$ which
might be a two-magnon bound state. 

A singlet bound-state with zero total momentum would be observable
by Raman scattering and would show up 
as a delta-function contribution
to the Raman intensity (\ref{I_new}):
$$
I_R(\omega)\ =\ A\delta(\omega-E_s)\ +\ I_R^\prime(\omega),
$$
with $I_R^\prime(\omega)$ beeing the continuum part of the
Raman intensity. A finite weight $A$ implies also a finite
value of the relative weight of the first pole,
\begin{equation}
A_r\ =\ \lim_{N_s\rightarrow\infty}{w_1\over\sum_n w_n},
\label{A}
\end{equation}
with $w_n=|\langle0|H_R|n\rangle|^2$.
In the inset of Fig.\ \ref{T=5} we show the
relative weight as a function of $1/N_s^2$ for
chains with up to $N_s=28$ sites and $\alpha=0.24$. 
We find no indication for an isolated bound state, 
i.e.\ for a finite value of $A_r$  both for $\delta=0$ (filled circles),
as expected, and for $\delta=0.03$ (filled triangles).
The data for $\delta=0.03$ shown in the
inset of Fig.\ \ref{T=5} have been calculated for
$\gamma=0.12$. The data for $\gamma=0$ and
$\delta=0.03$ are very similar.

We have therefore no indication from numerics for an isolated
singlet bound state, in agreement with our interpretation
of the 30cm$^{-1}$ peak as a continuum contribution truncated
by matrix-element effect, see Eq.\ (\ref{analytic}). Our
numerical results show, on the other hand, unambigously that
the lower edge of the Raman spectrum is pulled below the
non-interacting two-magnon density of states. This fact,
which is discussed in detail further below, indicates
a {\it continuum} of two-magnon bound states with zero total
momentum. This scenario is conceivable in view of the fact, 
that the two-magnon spectrum is made up out of four spinons

\begin{figure}
\caption{The calulated ratios of the $\Delta_R/\Delta_S$
         from chains with $N_s=24$ sites for various
         $\alpha$ and $\delta$. The shaded region indicates the
         experimentally observed values for CuGeO$_3$.
\label{ratio}
        }
\end{figure}
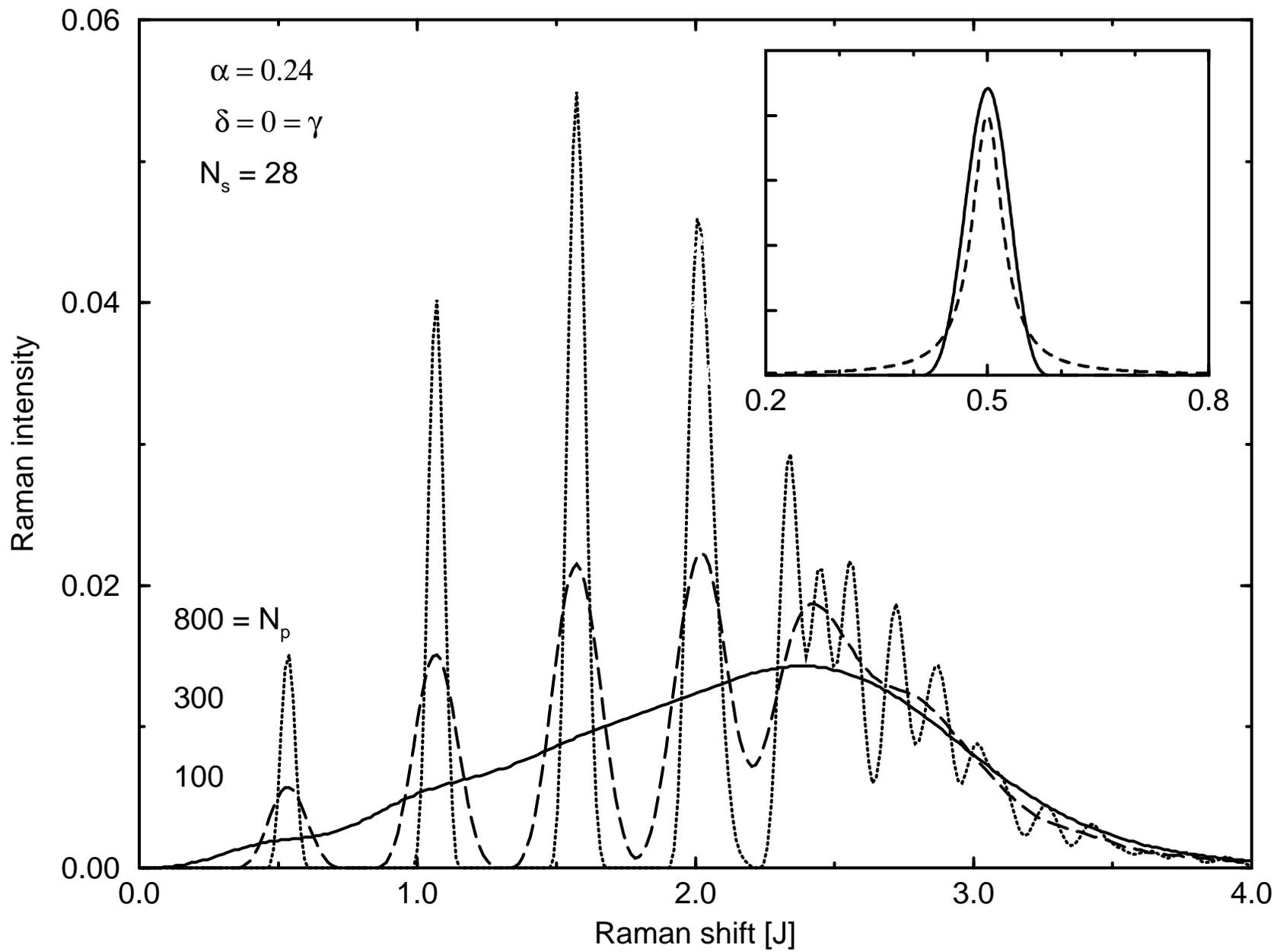
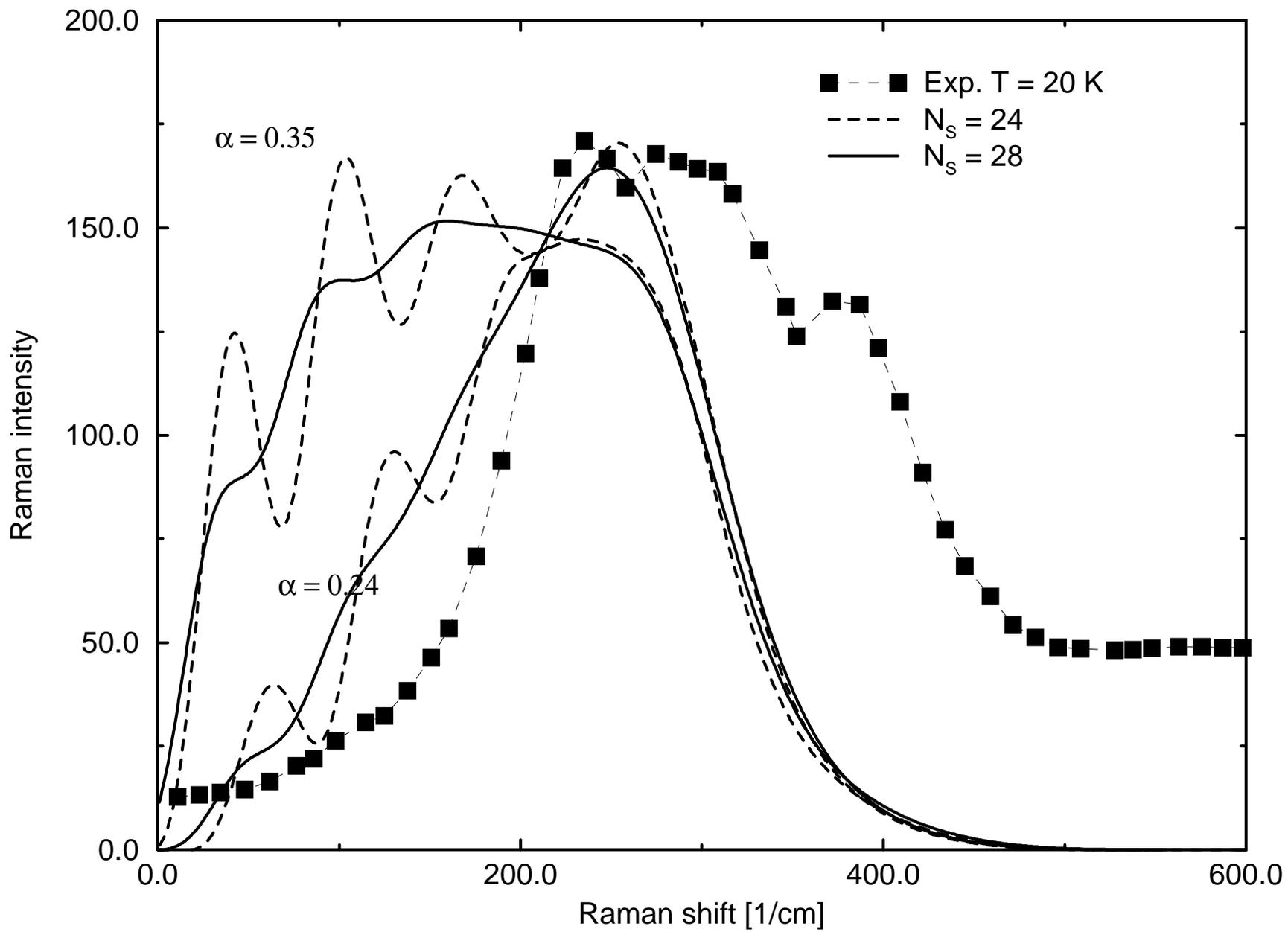
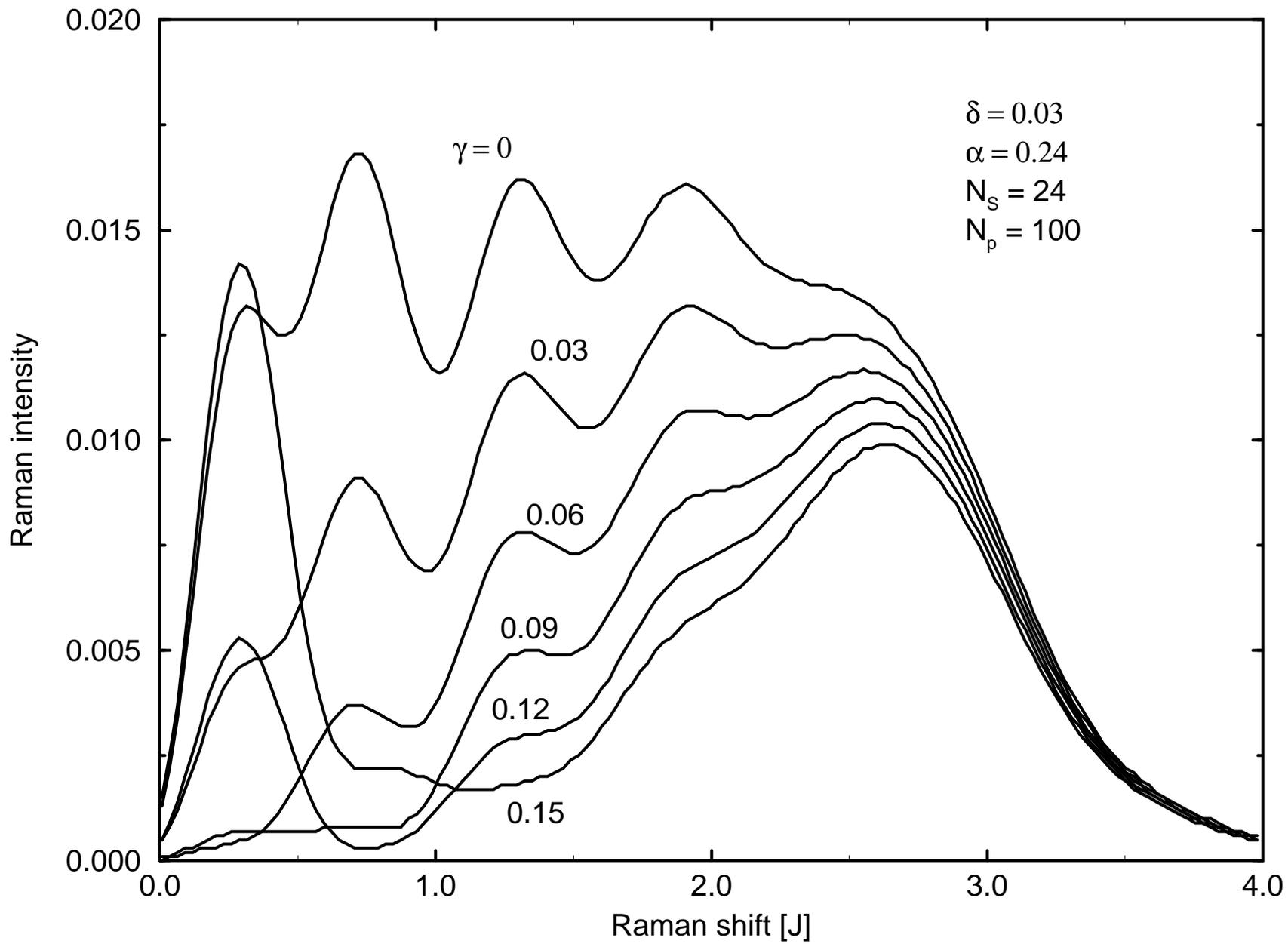
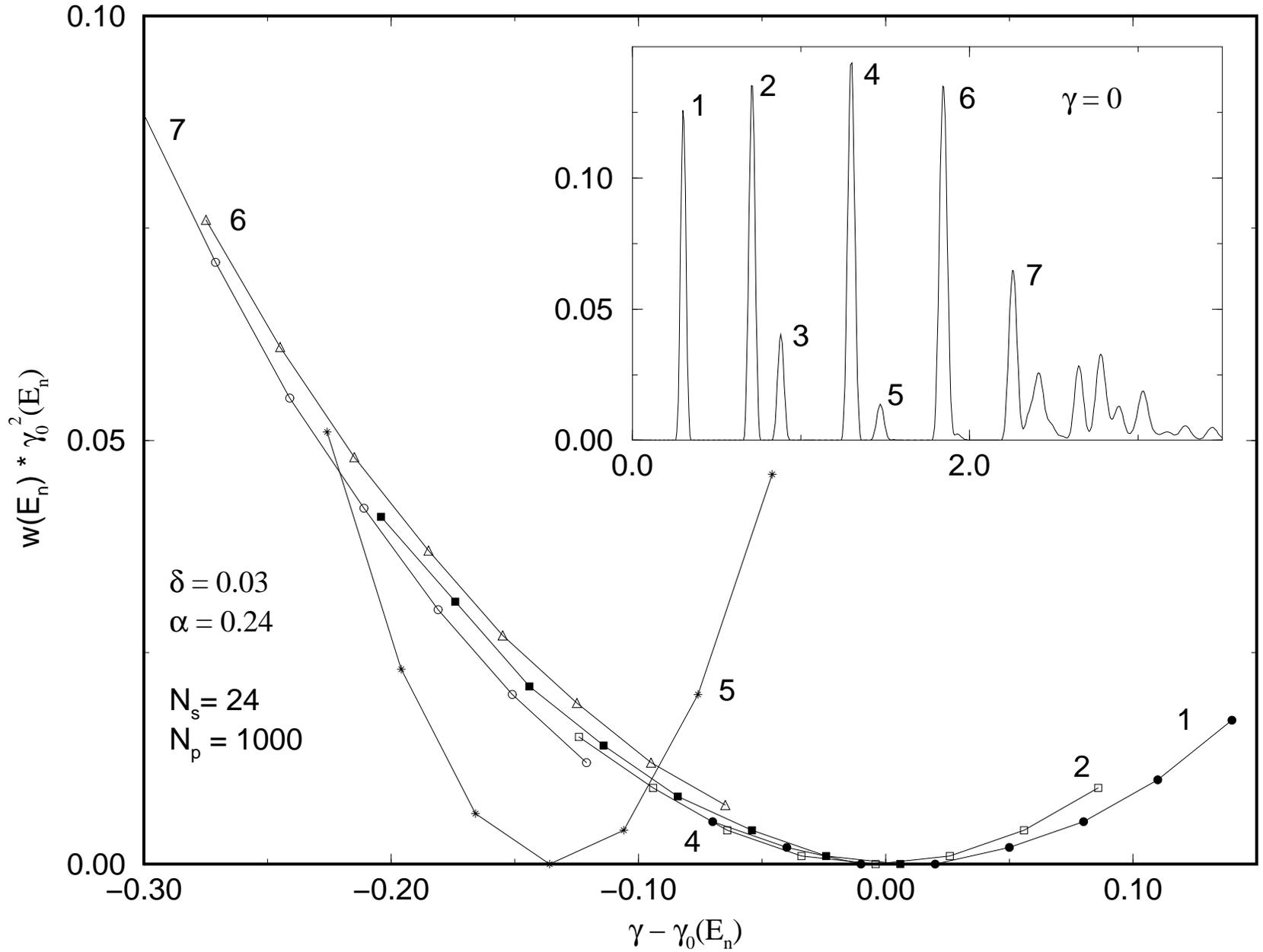

The experimentally observed values of single- and two-magnon
gap energies can be used, to a certain extent, to
determine the value of the parameters $J$, $\delta$ and
$\alpha$ entering Eq.\ (\ref{H}). The ratio
$\Delta_R/\Delta_S$ is independent of the coupling-constant
$J$. We have plotted $\Delta_R/\Delta_S$ 
in Fig.\ \ref{ratio}
for various values of $\alpha$ as
a function of $\delta$. The data are obtained for
$N_s=24$, the finite-size corrections are very
small (compare also \cite{Fledderjohann}). The experimental
possible values  1.49-1.78 are indicated as the shaded region
in Fig.\ \ref{ratio}. Again, we see that a larger value of
$\alpha\sim0.35$ does not do a good job on $\Delta_R/\Delta_S$.
In using Fig.\ \ref{ratio} for determing possible combinations
of $\alpha$ and $\delta$ one has to keep in mind that the
interchain coupling $J_b$, not included in our caculations,
has an effect of order 10\% on the values quoted in Fig.\ \ref{ratio}.
It is interesting to note, that recent experiments on
CuGeO$_3$ under pressure \cite{private} show that
the frustration parameter $\alpha$ increases with
presure, as does the experimental ratio
$\Delta_R/\Delta_S$, in accordance with the data
presented in Fig.\ \ref{ratio}.


In conlusion we have discussed in detail the effect of the
dimerization of the Raman operator on the Raman spectrum in
the spin-Peierls state of CuGeO$_3$. We have not found any
numerical evidence for a singlet bound-state and conclude
that the 30cm$^{-1}$ Raman line observed at $T=5K$ in
CuGeO$_3$ is a two-magnon line which is carved out of a
continuum by matrix-element effects due to the dimerization of
the Raman operator, $\gamma$. We have then compared the
observed ratio of the Raman gap to the neutron-scattering
gap and compared with numerical results. We conclude that this
ratio indicates an intermediate value for the frustration
parameter $\alpha$.

\acknowledgements
This work was supported through the Deutsche 
For\-schungs\-gemein\-schaft, 
the Graduierten\-kolleg ``Fest\-k\"{o}rper\-spekt\-roskopie'',
SFB 341 and SFB 252,
and by the BMBF 13N6586/8,

%
%
%
\begin{figure}
\caption{Dependence of the spectrum obtained by the
         kernel polyonmial approximation on the
         number of Tschebeycheff polynomials, $N_p=100,300,800$,
         retained.
         The data is for a chain with $N_s=28$
         sites and $\delta=0=\gamma$, $\alpha=24$.
         For $N_p\rightarrow\infty$ the poles evolve
         into delta-funtions. Insert: Comparison between
         a pole located at $x=0.5$ broaded by a Lorentzian
         $\epsilon/((x-0.5)^2+\epsilon^2)$ (dashed line), 
         with $\epsilon=0.025$ and an orthogonal polynomial 
         approximation with $N_p=100$ Tschebycheff polynomials.
\label{N_p}
             }
\end{figure}
\begin{figure}
\caption{The magnetic contribution to the experimental Raman 
         spectrum for CuGeO$_3$ at $T=20K$ (filled squares) and
         data obtained for chains with $N_s=24,28$
         sites (dashed/solid line) for $\alpha=0.24$
         and $\alpha=0.35$ respectively and $N_p=100$ and
         $\delta=0=\gamma$.
\label{T=20}
             }
\end{figure}
\begin{figure}
\caption{Dependence of the spectrum in the dimerized phase,
         $\delta=0.03$, $\alpha=0.24$, $N_p=100$ for a chain with $N_s=24$
         sites on the dimerization of the Raman-operator, 
         $\gamma=0,0.03,0.06,0.09,0.12,0.15$. Note the large
         matrix-element effects.
\label{gamma}
             }
\end{figure}
\begin{figure}
\caption{The rescaled weight $w(E_n)$ of the lowest seven poles
         in the Raman spectrum with pole Energy, $E_n$
         ($n=1,\dots,7$) as a function of the rescaled dimerization of 
         the Raman operator, $\gamma-\gamma_0(E_n)$, see 
         Eq.\ (\protect\ref{gamma_0}). The data is for a chain with
         $N_s=24$ sites and $\delta=0.03$, $\alpha=0.24$.
         The weight is rescaled by
         $\gamma_0(E_i)=\delta+\kappa E_i$, with 
         $\kappa\approx4/3000cm$. We observed that the
         rescaled weights of the dominant poles $1,2,4,6,7$
         fall all on a universal curve (the data for
         pole number 4 has been omitted in order to avoid
         overcrowding).
\label{weight_gamma}
             }
\end{figure}
\begin{figure}
\caption{Comparision between the experimental Raman
         spectrum in the dimerized phase of CuGeO$_3$
         at $T=5K$ (filled squares) and the analytic
         expression ((\protect\ref{analytic}), solid line)
         Insert: The relative weight (Eq.\ (\protect\ref{A}))
         of the lowest pole contribution to $I_R(\omega)$
         as a function of $1/N_s^2$ for $N_s=12,16,20,24,28$.
         The data is for $\alpha=0.24$, $\delta=0$ (filled circles)
         and $\delta=0.03$ and $\gamma=0.12$ (filled triangles).
\label{T=5}
        }
\end{figure}
\end{document}